\documentclass[11pt,urlcolor=blue, linkcolor=blue]{article} 
\usepackage{cite}

\usepackage{amsmath, amsthm, amssymb,slashed}
\usepackage{mathtools}

\usepackage{ifpdf}
\ifpdf
  \usepackage[pdftex]{graphicx}
  \usepackage{epstopdf}
\else
  \usepackage[dvips]{graphicx}
\fi
\usepackage[letterpaper]{geometry}
\parskip=\baselineskip

\usepackage{tablefootnote}

\usepackage[usenames, dvipsnames]{color}
\usepackage[svgnames]{xcolor}
\usepackage[colorlinks,citecolor=RoyalBlue, urlcolor=RoyalBlue, linkcolor=RoyalBlue ]{hyperref} 

\allowdisplaybreaks[1]
\numberwithin{equation}{section}

\newcommand{\cblue}[1]{\textcolor{blue}{#1}}
\newcommand{\cred}[1]{\textcolor{black}{#1}}

\usepackage{upgreek}

\usepackage{tabularx}

\DeclareMathAlphabet{\mathpzc}{OT1}{pzc}{m}{it}

\newcommand{\bea}{\begin{eqnarray}}
\newcommand{\eea}{\end{eqnarray}}
\def\be{\begin{equation}}
\def\ee{\end{equation}}

\def\re{{\mathrm{e}}}

\definecolor{red}{rgb}{1,0,0}
\definecolor{blue}{rgb}{0,0,1}
\definecolor{dblue}{rgb}{0,0,0.4}
\definecolor{green}{rgb}{0,1,0}
\definecolor{black}{rgb}{0,0,0}
\definecolor{white}{rgb}{1,1,1}

\definecolor{brn}{rgb}{.8,.4,.0}
\definecolor{redo}{rgb}{1,.5,.0}
\definecolor{ddgrn}{rgb}{0,0.4,0}
\definecolor{dgrn}{rgb}{0,0.55,0}
\definecolor{dbl}{rgb}{0,0,0.5}

\usepackage[bbgreekl]{mathbbol}
\usepackage{amscd}

\newcommand{\ii}{\hspace{1pt}\mathrm{i}\hspace{1pt}}
\newcommand{\dd}{\hspace{1pt}\mathrm{d}}

\newcommand{\Refe}[1]{Ref.~\cite{#1}}
\newcommand{\Eq}[1]{Eq.~(\ref{#1})} 
 
\newcommand{\eqn}[1]{Eq.~(\ref{#1})}

\newcommand{\prt}{\partial}

\newcommand{\bpm}{\begin{pmatrix}}
\newcommand{\epm}{\end{pmatrix}}
\newcommand{\bmm}{\begin{matrix}}
\newcommand{\emm}{\end{matrix}}

\newcommand{\cD}{ {\cal D} }

\def\CO{{\cal O}}
\def\CQ{{\cal Q}}

\newcommand{\Z}{\mathbb{Z}}

\newcommand{\R}{\mathbb{R}}

\def\Z{{\mathbb{Z}}}

\def\bZ{{\mathbf{Z}}}

\DeclareRobustCommand\sWang
{\cblue{\includegraphics[height=3.55ex]{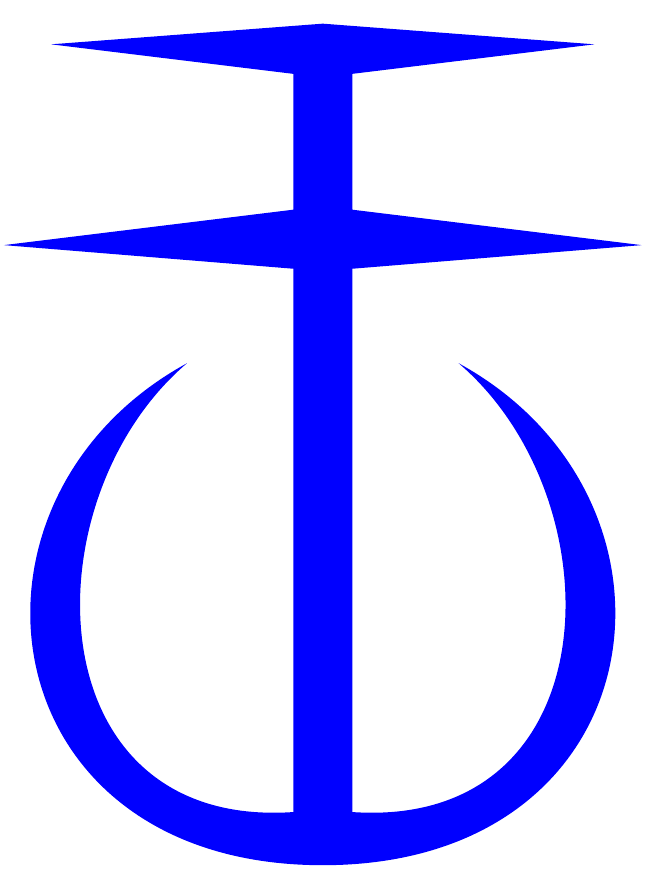}}}

\newcommand{\Wangfootnote}[1]{%
\let\oldthefootnote=\thefootnote%
\stepcounter{mpfootnote}%
\addtocounter{footnote}{-1}%
\renewcommand{\thefootnote}{\sWang}
\footnote{#1}%
\let\thefootnote=\oldthefootnote%
}

\DeclareRobustCommand\sXu%
{\cblue{\includegraphics[height=4.2ex]{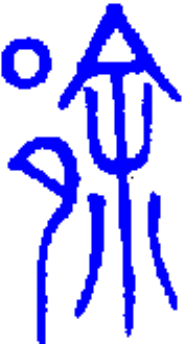}}}

\newcommand{\Xufootnote}[1]{%
\let\oldthefootnote=\thefootnote%
\stepcounter{mpfootnote}%
\addtocounter{footnote}{-1}%
\renewcommand{\thefootnote}{\sXu}
\footnote{#1}%
\let\thefootnote=\oldthefootnote%
}

\DeclareRobustCommand\sYau%
{\cblue{\includegraphics[height=3.4ex]{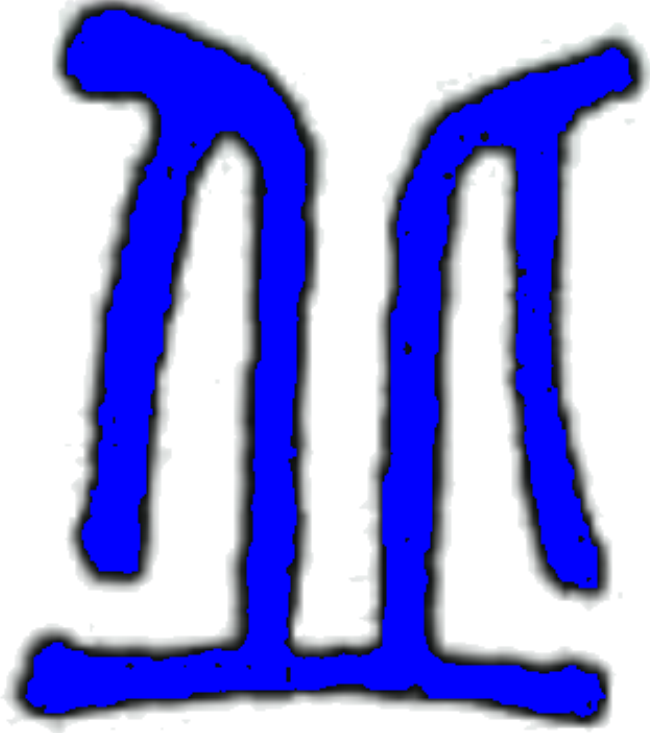}}}

\newcommand{\Yaufootnote}[1]{%
\let\oldthefootnote=\thefootnote%
\stepcounter{mpfootnote}%
\addtocounter{footnote}{-1}%
\renewcommand{\thefootnote}{\sYau}
\footnote{#1}%
\let\thefootnote=\oldthefootnote%
}

\usepackage{enumitem,cleveref}
\newcommand{\nn}{\nonumber}

\usepackage{comment}

\def \- {\!\smallsetminus\!}

\newcommand{\U}{{\rm U}}

\def \rH{\operatorname{H}}

\def \Z{\mathbb{Z}}

\newcommand{\Sec}[1]{Sec.~\ref{#1}}

\newcommand{\Fig}[1]{Fig.~\ref{#1}}

\usepackage{tikz}
\usetikzlibrary{calc}

\usepackage[all,cmtip]{xy}
\usepackage{tikz-cd}

\usetikzlibrary{matrix}
\usetikzlibrary{knots}

\usetikzlibrary{calc}

\usepackage[all,cmtip]{xy}
\usepackage{tikz-cd}

\usepackage{datetime}

\newenvironment{myfont}[2][]{\csname#2\endcsname[#1]}{}

\usepackage{makecell} 

\usepackage{slashed}
\usepackage[makeroom]{cancel}
\usepackage[normalem]{ulem} 
\usepackage{soul}

\usepackage[X2,T1]{fontenc}
\DeclareSymbolFont{cyrillic}{X2}{cmr}{m}{n}
\SetSymbolFont{cyrillic}{bold}{X2}{cmr}{bx}{n}
\DeclareMathSymbol{\khk}{\mathord}{cyrillic}{139}
\DeclareMathSymbol{\zh}{\mathord}{cyrillic}{117}
\DeclareMathSymbol{\zhen}{\mathord}{cyrillic}{182}
\DeclareMathSymbol{\zhe}{\mathord}{cyrillic}{134}

\def \U{\mathrm{U}}

\newcommand{\tm}{\hspace{1pt}\mathrm{m}}
\newcommand{\fD}{\hspace{1pt}\mathfrak{D}}
\newcommand{\tp}{\hspace{1pt}\mathrm{p}}

\newtheorem{defn}{Definition}

\newcommand{\Wpfootnote}[1]{%
\let\oldthefootnote=\thefootnote%
\stepcounter{mpfootnote}%
\addtocounter{footnote}{-1}%
\renewcommand{\thefootnote}{{{W\,}}}
\footnote{#1}%
\let\thefootnote=\oldthefootnote%
}

\newcommand{\Xfootnote}[1]{%
\let\oldthefootnote=\thefootnote%
\stepcounter{mpfootnote}%
\addtocounter{footnote}{-1}%
\renewcommand{\thefootnote}{{{X\;}}}
\footnote{#1}%
\let\thefootnote=\oldthefootnote%
}

\newcommand{\Yfootnote}[1]{%
\let\oldthefootnote=\thefootnote%
\stepcounter{mpfootnote}%
\addtocounter{footnote}{-1}%
\renewcommand{\thefootnote}{{{Y\;}}}
\footnote{#1}%
\let\thefootnote=\oldthefootnote%
}

\begin{document}
\begin{titlepage}
\begin{flushright}
\end{flushright}

\begin{center}

{\bf\LARGE{
Higher-Rank 
Tensor  Non-Abelian 
Field Theory:  \\[6.5mm]  
%
\Large{Higher-Moment or Subdimensional Polynomial Global Symmetry,}\\[6mm] 
Algebraic Variety, Noether's Theorem, and Gauging 
\\[5.5mm]  
}}

\vskip.5cm
\quad\quad\quad
\Large{
Juven Wang$^{1}$,\Wangfootnote{{e-mail: {\tt jw@cmsa.fas.harvard.edu
} (Corresponding Author) }} 
\quad    Kai Xu$^{2}$, \Xufootnote{{e-mail: {\tt  kaixu@math.harvard.edu}}}
and  Shing-Tung Yau$^{1,2,3}$ \Yaufootnote{ {e-mail: {\tt  yau@math.harvard.edu}
\hfill  October 2019 \\[2mm]
{
  \emph{Dedicated to 
90 years of Gauge Principle since Hermann Weyl} [Elektron und Gravitation, Zeit. f\"ur Physik 56, 330-352 (1929)]
}
} 
}
}
 
\vskip.5cm
{\small{\textit{$^1${Center of Mathematical Sciences and Applications, Harvard University,  Cambridge, MA 02138, USA} \\}}
}
\vskip.2cm
{\small{\textit{$^2$ Department of Mathematics, Harvard University, Cambridge, MA 02138, USA}\\}}
\vskip.2cm
{\small{\textit{$^3$ Department of Physics, Harvard University, Cambridge, MA 02138, USA}\\}}

\end{center}
\vskip.35cm
\baselineskip 12pt
\begin{abstract}

With a view toward a fracton theory  
in condensed matter,
we introduce a higher-moment polynomial \cred{degree-p} 
global symmetry, acting on complex scalar/vector/tensor fields
\cred{(e.g., ordinary or vector global symmetry for p$=0$ and p$=1$ respectively)}.
We relate this higher-moment global symmetry of $n$-dimensional space, 
to a lower degree (either ordinary or higher-moment, e.g., degree-(p-$\ell$)) 
subdimensional or subsystem global symmetry on layers of $(n-\ell)$-submanifolds.
These submanifolds are algebraic affine varieties (i.e., solutions of polynomials). 
The structure of layers of submanifolds as subvarieties can be studied via mathematical tools of embedding, foliation, and algebraic geometry.
We also generalize Noether's theorem for this higher-moment polynomial global symmetry.
We can promote the higher-moment global symmetry to a local symmetry,
and derive a new family of higher-rank-m symmetric tensor gauge theory by gauging, \cred{with m = p$+1$}.
By further gauging a discrete \cred{$\mathbb{Z}_2^C$ charge conjugation (particle-hole)} symmetry,
we derive a new general class of  rank-m 
tensor non-abelian gauge field theory (the gauge structure is non-commutative thus non-abelian but not an ordinary group): a hybrid class of (symmetric or non-symmetric) higher-rank-m tensor gauge theory 
and anti-symmetric tensor topological field theory, 
generalizing [arXiv:1909.13879], interplaying between gapless and gapped sectors.\\[12mm]


\end{abstract}
\end{titlepage}


\renewcommand{\eqref}{\eqn}

  \pagenumbering{arabic}
    \setcounter{page}{2}
    
\tableofcontents   



\section{Introduction
}
\label{sec:intro}

\cred{Fracton orders \cite{RahulNandkishore2018sel1803.11196, PretkoReview2020cko2001.01722}
are the new kinds of orders in many-body quantum matter systems. Fracton orders are defined physically by exhibiting
some of (if not all of) the following properties:\footnote{We should focus on the limited references essential to the construction of our theories. We pardon for potentially leaving out some other important works from the References. For more References in condensed matter literature, the readers can find in 
the review \Refe{RahulNandkishore2018sel1803.11196, PretkoReview2020cko2001.01722}. For a short historical account of gauge theory and the earlier References, 
starting from the basics of Maxwell electromagnetism and Weyl gauge principle \cite{Weyl1929fm}, the readers can find in \cite{Wang2019aiq1909.13879}.}
\begin{itemize}
\item For gapped fractons, their ground state degeneracy (GSD) is similar to topological order \cite{1610.03911} with GSD depending on the topology of base space or spatial manifolds.  
Moreover, they have extensive GSD depending on the system size and the details of lattice sites and cutoffs \cite{2005PRL0404182Chamon, 2011PhRvAHaah1101.1962}. 
\item Fracton orders can also
have excitations, either being immobile in isolation or being mobile moving along in lower dimensions or subdimensions \cite{Vijay2015mka1505.02576VijayHaahFu, Vijay2016phm1603.04442}.
\item Fracton orders are associated with the long-range entangled phases of quantum matter, obtainable by dynamically gauging the \emph{subsystem global symmetries} 
or \emph{subdimensional global symmetries} of the full quantum systems \cite{Vijay2016phm1603.04442, Williamson1603.05182, XieChen1806.08679}
(see also earlier work \cite{SavvidyWegner9308094, Savvidy9311026} before the fracton concept is introduced).
\end{itemize}}

Motivated by the fracton order in condensed matter,
recently two of the present authors introduced a new hybrid family of tensor gauge field theories \cite{Wang2019aiq1909.13879}
mixing between the anti-symmetric tensor gauge fields and symmetric higher-rank tensor gauge fields in a delicate way.
Their purpose was to formulate the first toy model of a gauge theory with a non-abelian continuous gauge structure for fracton order in condensed matter (see reviews\cite{RahulNandkishore2018sel1803.11196, PretkoReview2020cko2001.01722}).
\cred{The toy model  \cite{Wang2019aiq1909.13879} suggested an interplay between: 
\begin{enumerate}[leftmargin=2.0mm, label=\textcolor{blue}{\arabic*}., ref=\textcolor{blue}{\arabic*}]
\item 
the gapped anti-symmetric tensor gauge topological quantum field theory (TQFT)
with topological order, and 
\item the gapless symmetric higher-rank tensor gauge theory with \emph{gapless higher-spin U(1)-gauge photon-like modes}.
\end{enumerate}
}
The higher-rank tensor gauge theory in \Refe{Wang2019aiq1909.13879} combines the feature of: 
\\[-10mm]
\begin{enumerate}[leftmargin=2.0mm, label=\textcolor{blue}{\arabic*}., ref=\textcolor{blue}{\arabic*}]
\item \cred{Anti-symmetric tensor topological field theory (TQFT):
We adopt a continuum TQFT formulation of \label{model-asym}
group-cohomology topological gauge theory (known as Dijkgraaf-Witten theory or twisted gauge theory \cite{DijkgraafWitten1989pz1990}) by  anti-symmetric tensor
differential form gauge fields (i.e., Kalb-Ramond fields \cite{KalbRamond1974yc}).}
There in \Refe{Wang2019aiq1909.13879} and here, we mainly use a particular continuum TQFT formalism set-up and notations presented in \cite{1602.05951WWY, Putrov2016qdo1612.09298, Wang2018edf1801.05416, Wang1901.11537WWY}
that can capture all finite \emph{abelian} unitary gauge group and some \emph{non-abelian} unitary gauge group of Dijkgraaf-Witten theory with the group-cohomology cocycle twist.
(See also other related general formulations for non-dynamical gauge background theories \cite{Kapustin1404.3230, Wang1405.7689} and references therein.)

\item Symmetric tensor field theory: \label{model-sym}
 We will only apply a specific class of \emph{symmetric higher-rank tensor gauge theories}
or \emph{higher-spin theories} studied in the condensed matter literature, e.g.
\Refe{2016arXiv160108235RRasmussenYouXu, Pretko2016kxt1604.05329, Pretko2016lgv1606.08857, Pretko2017xar1707.03838, Slagle2018kqf1807.00827, Pretko2018jbi1807.11479}, largely inspired by Pretko's work.
\end{enumerate}
\Refe{Wang2019aiq1909.13879} finds that a class of symmetric higher-rank tensor gauge field theory (from the Model \ref{model-sym} of the symmetric tensor field theory)
by gauging a higher-moment abelian vector global symmetry
U(1)$_{x_{(n)}}$\footnote{Here we denote a vector global symmetry along $n$-dimensions as $ \U(1)_{x_{(n)}}$: 
The $ \U(1)_{x_{(d+1)}}$ means the vector global symmetry in a $d+1$-dimensional spacetime,
and the $ \U(1)_{x_{(d)}}$ means the vector global symmetry in a $d$-dimensional \cred{space}.
The $x$ means to be a set of Cartesian coordinates for the spacetime. \\
Throughout this article,
we should only focus on a flat Euclidean or Minkowski spacetime ($\R^{d+1}$ or $\R^{1,d}$) with Cartesian coordinates. The Cartesian coordinates can be easily realized
in a square, rectangular, cubic lattices in condensed matter systems. On the other hand, it is difficult to imagine
how to rewrite the higher-moment global symmetry in curved coordinates such as spherical or cylindrical coordinates, and how to realize them 
in a lattice system with energy cutoffs in condensed matter.\\
{The $d+1$d means the $d+1$ spacetime dimensions, with $d$ spatial and 1 time dimensions.
The $D$d means the $D$ spacetime dimensions.
The $\bar{D}$D means the $\bar{D}$ space dimensions.
We denote $d+1$D means the $d$ spatial and 1 time dimensions.}} 
and also an ordinary 0-form global symmetry known as a $\Z_2^C$-charge conjugation \cred{(particle-hole)} symmetry.
The higher-moment symmetry U(1)$_{x_{(n)}}$ and the $\Z_2^C$-charge conjugation symmetry do \emph{not} commute. 
As we shall elaborate below,
they form a semi-direct product (denoted $\ltimes$) structure.
\Refe{Wang2019aiq1909.13879} dynamically gauges the  $\Z_2^C$-charge conjugation symmetry to gain a non-abelian gauge structure:\footnote{This $\Z_2^C \ltimes \Big(  \U(1)_{x_{(n)}} \Big)$ 
is \emph{not} quite an ordinary group structure. However, because the polynomial symmetry operation $\U(1)_{x_{(n)}}$ does not commute with the 
$\Z_2^C$ symmetry, we still stick to the standard convention to call the non-commutative structure as a non-abelian structure.
We also call the non-commutative gauge structure (although not a gauge group) a 
non-abelian gauge structure.
}
\bea \label{eq:Z2CU1xn-gauge}
\left[\Z_2^C \ltimes \Big(  \U(1)_{x_{(n)}} \Big) \right]. 
\eea
This gauge structure is the first example in the fracton order literature satisfying the properties below:\footnote{We denote the global symmetry in the bracket $[...]$ to imply that it is dynamically gauged.} 
\begin{itemize}
\item compact,
\item continuous,\footnote{We should remind the readers that there are alternative pursuits to construct nonabelian fracton orders with discrete gauge structures on a lattice\cite{Vijay2017cti1706.07070, SongPremHuang2018gbb1805.06899, PremHuangSong2018jsn1806.04687, BulmashMaissamBarkeshli2019taq1905.05771, PremWilliamson2019etl1905.06309} 
(e.g. discrete gauge theories) instead of continuous gauge structures. It is possible to Higgs down our model with continuous gauge structures to 
obtain higher-rank nonabelian tensor field theories with nonabelian discrete gauge structures \cite{WXY4}.}
\item non-abelian (gauging $\Z_2^C$) or abelian (not gauging $\Z_2^C$),
\item with two disconnected pieces in the gauge structure due to $\Z_2^C$. 
\end{itemize}
The $ \U(1)_{x_{(n)}}$ actually means there are $n$-independent vector directions $x_i$ with $i=1, \dots, n$. It is easier to understand 
$ \U(1)_{x_{(n)}}$ before gauging it. Thus, let us first recover the gauge structure \Eq{eq:Z2CU1xn-gauge} to the ungauged global symmetry:
\bea \label{eq:Z2CU1xn-global}
\Z_2^C \ltimes \Big( \U(1) \times \U(1)_{x_{(n)}}\Big). 
\eea
We can perform an ordinary 0-form global symmetry U(1) and a vector global symmetry $ \U(1)_{x_{(n)}}$ 
transformations
by transforming a complex matter field $\Phi \in \mathbb{C}$ (following the pioneer work of Pretko's \cite{Pretko2018jbi1807.11479} and \cite{Wang2019aiq1909.13879}) 
\bea \label{eq:vglobal-1}
\Phi \to \re^{\ii Q(x)}\Phi= \re^{\ii (\Lambda_{i}  x_i + \Lambda_0)}\Phi.
\eea 
\begin{itemize}
\item
When the $\Lambda_{i}$ (as a constant, independent of spacetime coordinates) is nonzero, we have a degree-1 polynomial $Q(x)=(\Lambda_{i}  x_i + \Lambda_0)$ of $x$ 
which  $\Lambda_{i}  x_i $ specifies the 
$\U(1)_{x_{(n)}}$ vector global symmetry,
while $\Lambda_{0}$ specifies the  $\U(1)$ ordinary global symmetry. 
Since there are n independent $\Lambda_{i}  x_i$,
we have indeed several copies of commuting $\U(1)_{x_j}$-vector global symmetry:
\bea \label{eq:U1-degree-1}
\U(1)_{x_{(n)}}:= \U(1)_{x_{1}}\times \U(1)_{x_{2}}\times\dots \times \U(1)_{x_{n}}=\prod_{j=1}^n \U(1)_{x_{j}}.
\eea
We name this global symmetry as a degree-1 global symmetry, thanks to the degree-1 polynomial $Q(x)=(\Lambda_{i}  x_i + \Lambda_0)$.
\item
When the $\Lambda_{i}$ is zero,  we have a degree-0 polynomial $(\Lambda_0)$ independent of $x$. 
We name this global symmetry as a degree-0 global symmetry thanks to the degree-0 polynomial $Q(x)=\Lambda_0$.
\end{itemize}
If we gauge only the vector global symmetry $\U(1)_{x_{(n)}}$ but not the ordinary symmetry U(1),
we gain the $[\U(1)_{x_{(n)}}]$ gauge structure, while the remained U(1) is neither gauged nor global symmetry anymore.
This particular way of gauging $[\U(1)_{x_{(n)}}]$ introduces the compact symmetric rank-2 tensor gauge field $A_{ij}$.
%
\Refe{Wang2019aiq1909.13879} dynamically gauge the discrete $\Z_2^C$ symmetry which flips
$$
A_{ij} \to - A_{ij}
$$
to gain the non-abelian gauge structure.

\begin{figure}[!h]
	\centering
	\includegraphics[width=10.cm]{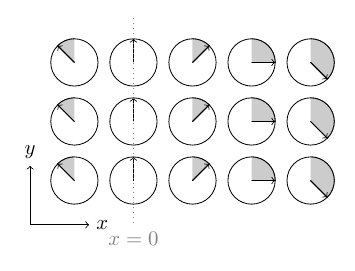}
	\caption[caption]{\cred{The {vector global symmetry} \Eq{eq:vglobal-1} belongs to a generalized class of higher-moment symmetry \Eq{eq:U1-poly-transf}. 
	For demonstration, here we show various $\Phi(x)$ fields sitting on discretized lattice points on the $(x,y)$ plane.
	The {vector} U(1) global symmetry transformation acts on the complex charged matter $\Phi(x) \in \mathbb{C}$ (the rotor fields) as \Eq{eq:vglobal-1}:
$\Phi \to e^{\ii  Q( x)} \Phi:= e^{\ii  \Lambda \cdot x} \Phi$.
The angle $\Lambda \cdot x$ depends on a reference point (say $x=0$) and the distance $x$ away from the reference point.
	The clockwise angle (drawn in gray area) away from the 12 o'clock direction implies the complex phase of $\Phi(x) \in \mathbb{C}$.
}
	}
	\label{Fig1-1911}
\end{figure}

\begin{figure}[!h]
	\centering
		\includegraphics[width=16.cm]{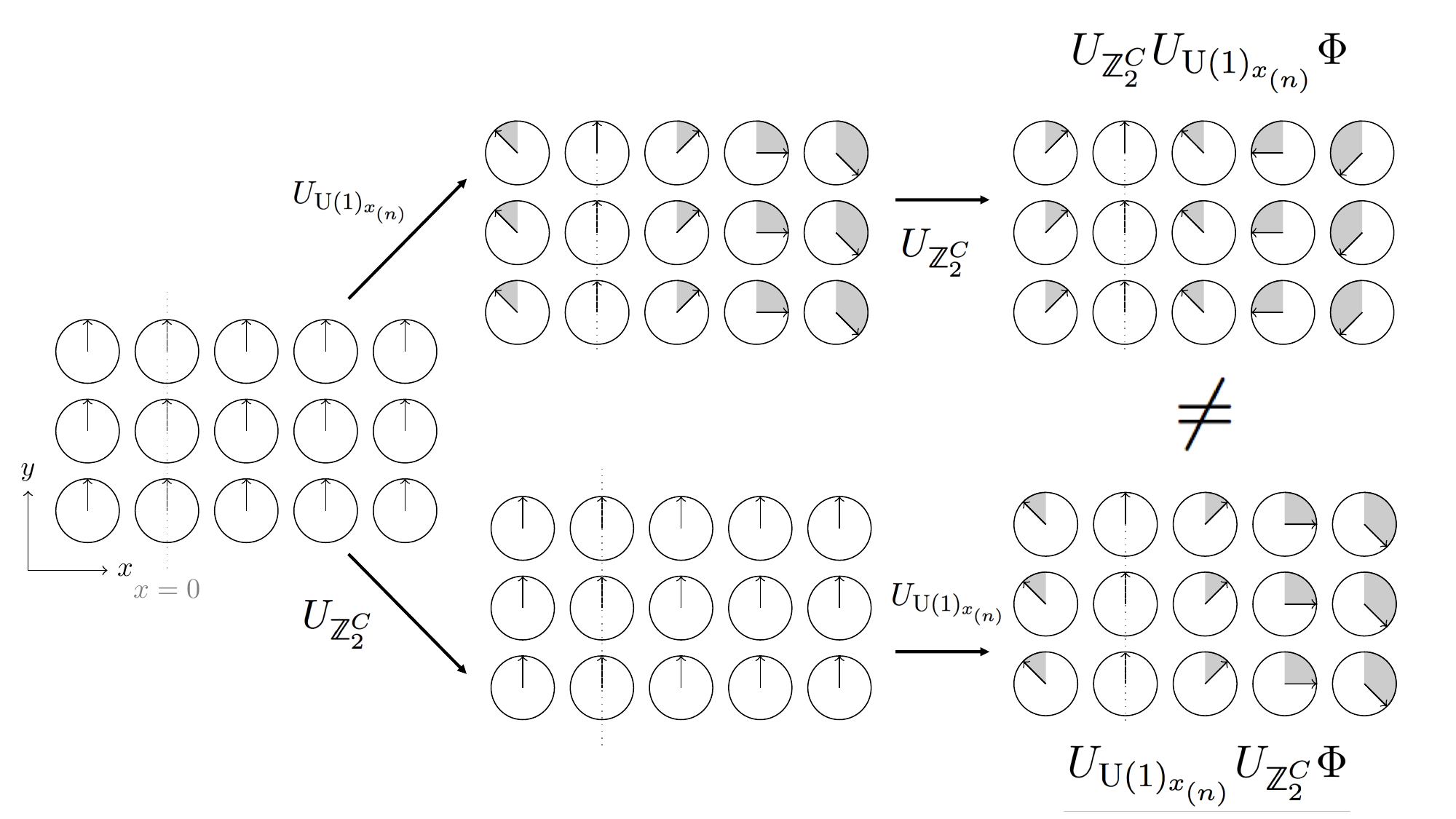}
		\caption[caption]{\cred{Follow the setup of \Fig{Fig1-1911}, here we show the non-commutative nature of
		the $\Z_2^C \ltimes   \U(1)_{x_{(n)}}$ global symmetry by performing the $U_{\Z_2^C} U_{{\rm{U}}(1)_{x_{(n)}}}  \Phi$ transformation
		on the top route,
		versus the $U_{{\rm{U}}(1)_{x_{(n)}}} U_{\Z_2^C} \Phi$ transformation
	        on the bottom route.
		\quad
The top route shows 
$U_{\Z_2^C} U_{{\rm{U}}(1)_{x_{(n)}}}  \Phi =U_{\Z_2^C} ( e^{\ii  Q( x)} \Phi)
=e^{\ii  Q( x)} \Phi^\dagger.$
The bottom route shows 
$U_{{\rm{U}}(1)_{x_{(n)}}} U_{\Z_2^C} \Phi =U_{{\rm{U}}(1)_{x_{(n)}}} (\Phi^\dagger)
=e^{-{\ii  Q( x)}} \Phi^\dagger$.
In summary, we demonstrate
$U_{\Z_2^C} U_{{\rm{U}}(1)_{x_{(n)}}}  \Phi \neq U_{{\rm{U}}(1)_{x_{(n)}}} U_{\Z_2^C} \Phi$.
Indeed the non-commutative nature between the $\Z_2^C$ and the higher moment global symmetry
still holds, even for a more general polynomial global symmetry
(such as \Eq{eq:U1-poly-transf}'s transformation $\Phi\to\re^{\ii Q(x)}\Phi := \re^{\ii \big(\Lambda_{i_1,\dots,i_{\tm-1}}  x_{i_1} \dots x_{i_{\tm-1}} +\dots+ \Lambda_{i,j}  x_ix_j + \Lambda_{i}  x_i + \Lambda_0\big)}\Phi$). 
	}}
	\label{Fig2-1911}
\end{figure}

\cred{To help the readers digesting the non-abelian gauge structure 
$\left[\Z_2^C \ltimes \Big(  \U(1)_{x_{(n)}} \Big) \right]$ in \Eq{eq:Z2CU1xn-gauge},
here we show the non-commutative symmetry operations between the
$\U(1)_{x_{(n)}}$ vector global symmetry transformation (\Fig{Fig1-1911}) and 
the $\Z_2^C$
charge conjugation (particle-hole) symmetry transformation 
on a complex bosonic scalar field $\Phi(x) \in \mathbb{C}$
in cartoon figures, see \Fig{Fig2-1911}.}

\begin{figure}[!h]
	\centering
	\includegraphics[width=11.cm]{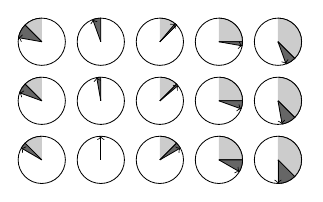}
	\caption[caption]{\cred{A demonstration of the local fluctuations (the dark gray color) deviate away from the 
 {vector (degree-1 polynomial) global symmetry  of  \Eq{eq:vglobal-1}
 or higher moment polynomial global symmetry} of \Eq{eq:U1-poly-transf}. The light gray color indicates higher-moment degree-(m-1) global symmetry transformations (here we show m $=1$).
 We will introduce a gauge theory  
 to compensate 
 the local gauge fluctuation  in \Sec{sec:ScalarChargeHigher-MomentPolynomial}
 by introducing a rank-m tensor gauge field $A_{i_1,\cdots,i_{\tm}}$ (\Eq{eq:abelian-gauge-transformation-sym-rank-m-A}) mediating between the matter $\Phi$ fields.}
	}
	\label{Fig3-1911}
\end{figure}

\newpage
In this work, we follow the setup in \cite{Wang2019aiq1909.13879}, and proceed to develop other families of new theories.
We consider the following generalization:
\begin{enumerate}[leftmargin=2.0mm, label=\textcolor{blue}{(\arabic*)}., ref=\textcolor{blue}{(\arabic*)}]
\item Higher-moment global symmetry 
for a complex scalar charge field $\Phi \in \mathbb{C}$:
A general polynomial degree-$(\tm-1)$ global symmetry\footnote{\cred{Throughout our work, 
we define $(\tm-1) \equiv \tp$ for some integers $\tm \geq 1$ and $\tp \geq 0$. The reason to
 choose $\tm \geq 1$ will become clear later when we require to construct the Lagrangian from
 the m-th derivative term $\prt^{\tm} \log \Phi$.}} 
on  $\Phi $ allows a symmetry transformation
\bea \label{eq:U1-poly-transf}
\Phi\to\re^{\ii Q(x)}\Phi := \re^{\ii \big(\Lambda_{i_1,\dots,i_{\tm-1}}  x_{i_1} \dots x_{i_{\tm-1}} +\dots+ \Lambda_{i,j}  x_ix_j + \Lambda_{i}  x_i + \Lambda_0\big)}\Phi.
\eea
Here all
$\Lambda_{\dots}$ in \Eq{eq:U1-poly-transf} are constants independent of spacetime coordinates. 
When we gauge such a higher-moment polynomial degree-$(\tm-1)$ global symmetry,
we will introduce a rank-m compact symmetric tensor gauge field $ A_{i_1,\dots,i_{\tm}}$.
%
This new result is done in \Sec{sec:ScalarChargeHigher-MomentPolynomial}.

Let us generalize the notation of the symmetry in \Eq{eq:U1-degree-1} into
\bea
{\U(1)_{x_{n \choose M}^{M}}:= \prod_{\{j_1,\dots,j_{M}\}} \U(1)_{x_{j_1},\dots,x_{j_{M}}},} 
\eea
 for \emph{each} \cred{of} different $\Lambda_{j_1,\dots,j_{M}}$ independent coefficients of the symmetry generators from a degree $M$ polynomial.
 Therefore, for the full \Eq{eq:U1-poly-transf},  if every coefficient $\Lambda$ is allowed, then 
by combining with all the symmetry generators from degree 0, 1, $\dots$ to degree  ${\tm-1}$, we have the full symmetry structure: 
\bea \label{eq:U1-polynomial-product-group}
    \U(1)_{x_{ n \choose {\tm-1}}^{{\tm-1}}} \times
 \dots
 \times \U(1)_{x_{(n)}} \times \U(1)    
= 
\prod_{M=0}^{\tm-1} \U(1)_{x_{n \choose M}^{M}}.
\eea

\item Higher-moment global symmetry for a complex vector charge field $\Phi_I \in \mathbb{C}$ (where the vector index is $I$, but each $\Phi_I$ still is a complex scalar field):
A general polynomial degree-$({\tm}-1)$ global symmetry on  $\Phi_I$  allows a symmetry transformation
\bea 
\Phi_I\to\re^{\ii Q_I(x)}\Phi_I := \re^{\ii \big(\Lambda_{I;i_1,\dots,i_{{\tm}-1}}  x_{i_1} \dots x_{i_{\tm}-1} +\dots+ \Lambda_{I; i,j}  x_ix_j + \Lambda_{I; i}  x_i + \Lambda_{I;  0}\big)}\Phi_I.
\eea
The degree-$({\tm}-1)$ polynomial $Q_I(x) = (\Lambda_{I;i_1,\dots,i_{\tm}-1}  x_{i_1} \dots x_{i_{\tm}-1} +\dots+ \Lambda_{I; i,j}  x_ix_j + \Lambda_{I; i}  x_i + \Lambda_{I;  0}\big)$
has an index $I$.

\item
Higher-moment global symmetry for a complex rank-$M$ 
tensor charge field $\Phi_{I_1, \dots, I_M} \in \mathbb{C}$ (where the tensor index is ${I_1, \dots, I_M}$, but each $\Phi_{I_1, \dots, I_M}$ still is a complex scalar field):
A general polynomial degree-$({\tm}-1)$ global symmetry on  $\Phi_{I_1, \dots, I_M}$  allows a symmetry transformation
	\bea
	\Phi_{I_1, \dots, I_M}\to \re^{\ii Q_{I_1, \dots ,I_M}(x)} \Phi_{I_1, \dots, I_M}.
	\eea
\item For all the above theories of higher-moment global symmetries, we construct their corresponding gauge theories by dynamically gauging the global symmetry.
We introduce the abelian and non-abelian tensor gauge field, and their gauge invariant or covariant field strength tensor.
We can use the field strength to construct the gauge invariant kinetic Lagrangian term of the non-abelian tensor gauge theory, shown in 
\Sec{sec:AbelianF} and \Sec{sec:NAbelianF}.
\end{enumerate}

We also notice that \Refe{GriffinPetrHorava2014bta1412.1046, Gromov2018nbv1812.05104}
had also attempted to study 
the polynomial types or higher-moment types of global symmetries systematically.
Although our motivations are
somehow different from \cite{GriffinPetrHorava2014bta1412.1046} and our framework is somehow different from \cite{Gromov2018nbv1812.05104}.
We do not yet know a precise correspondence between our results \cite{Wang2019aiq1909.13879, WXY3Wang2019mtt1912.13485, WXY4} and theirs \cite{GriffinPetrHorava2014bta1412.1046, Gromov2018nbv1812.05104}.

Our theory can be formulated compatible with or without Euclidean, Poincar\'e, isotropic or anisotropic symmetry in the $d+1$d spacetime, 
at least in ultraviolet high or intermediate energy field theory, but not yet to a lattice cutoff scale, see more discussions in various versions of theories in \cite{Wang2019aiq1909.13879}.
Thus for Euclidean or Poincar\'e symmetry, we need to choose the $n$-dimensions
in $ \U(1)_{x_{(n)}}$ as $n= d+1$ for dimensions. 
For an anisotropic symmetry, we can 
choose the $n$-dimensions
in $ \U(1)_{x_{(n)}}$ as $n \leq d$ for dimensions. 
Below we shall keep the general index $n$ in
$ \U(1)_{x_{(n)}}$, and leave the substitution of $n$ freely (to $n= d+1$ or  $n \leq d$) based on the specific needs of readers. 


In a companion work, we explore the new types of sigma model that can interpolate between the disorder phases (as the present higher-rank tensor non-abelian gauge theories)
and the ordered phases. Similar to the famous quantum phase transition between insulator 
(U(1) symmetry disorder described by a topological gauge theory or a disordered Sigma model)  
and superfluid/superconductivity (U(1) global/gauge symmetry-breaking order described by a Sigma model with a U(1) target space with Goldstone modes),
we can explore phase structures of  order-disorder phases by developing a new Sigma model \cite{WXY3Wang2019mtt1912.13485}.
 A recent work studies the superfluid phase of a pure abelian fractonic matte field theory without gauge fields \cite{Yuan2019gehPengYe1911.02876},
 while we study instead both the ordered phase (superfluid and a sigma model) 
 and the disordered phase of non-abelian gauged fractonic matte field theories with gauge fields \cite{WXY3Wang2019mtt1912.13485}.

 

\section{Scalar Charge, Higher-Moment Polynomial Degree-(m-1) Global Symmetry, and Rank-m Gauge Theory} 
\label{sec:ScalarChargeHigher-MomentPolynomial}

	First we consider how to gauge the following global symmetry for s scalar field $\Phi$ on the $\mathbb{R}^n$, the $n$-dimensional space or spacetime:
	\be \label{eq:degree-sym}
	\Phi\to\re^{\ii Q(x)}\Phi
	\ee
	where $Q(x)$ is a polynomial with degree at most $(\tm-1) \equiv \tp$ for some integers $\tm \geq 1$ and $\tp \geq 0$, say
	\be \label{eq:Qx-general}
Q(x) :=	(\Lambda_{i_1,\dots,i_{\tm-1}}  x_{i_1} \dots x_{i_{\tm-1}} +\dots+ \Lambda_{i,j}  x_ix_j + \Lambda_{i}  x_i + \Lambda_0\big).
	\ee
Note that the gauge transformation can be written as:
	\be
	\log \Phi \to \log\Phi+ \ii Q(x)
	\ee
and the only invariant quantity under this transformation is 
\bea
\prt^{\tm}\log\Phi,
\eea 
which is an order m symmetric tensor whose components are 
\bea
\prt_{i_1}\cdots\prt_{i_{\tm}}\log\Phi,
\eea 
where the spacetime indices $i_k\in \{1,2,\cdots, n\}$, with $k \in  \{1,2,\cdots, \tm\}$.
In the next subsection,  
\emph{before} gauging this higher-moment symmetry,
we construct the \emph{covariant operator} ($P_{i_1,\cdots,i_m}$ in \Eq{eq:covariant-P}).
\emph{After} gauging this higher-moment symmetry,
 we also construct the 
 \emph{gauge invariant operator} (such 
 as the {Abelian gauge field strength} in \Sec{sec:AbelianF})
 or  \emph{gauge covariant operator} (such 
 as  the covariant derivative on the matter field $D_{i_1,\cdots,i_m}[\{\Phi \}]$ in \Eq{eq:gauge-covariant-D-P-A}
 or the {non-Abelian gauge field strength} in \Sec{sec:NAbelianF}).

\subsection{Polynomial with arbitrary degree}
	\label{sec:degree-m}

	By the law of differentiation, we know that 
	\bea
	\prt_{i_1}\cdots\prt_{i_{\tm}}\log\Phi=\frac{P_{i_1,\cdots,i_{\tm}}(\Phi,\cdots,\prt^{\tm}\Phi)}{\Phi^{\tm}},
	\eea 
	so this order m tensor $P$ transforms as 
	\be \label{eq:covariant-P}
	P_{i_1,\cdots,i_{\tm}}\to\re^{\ii {\tm} Q(x)} P_{i_1,\cdots,i_{\tm}}.
	\ee
Under a more general gauge transformation 
	\be
	\Phi\to\re^{\ii \eta (x)}\Phi,
	\ee
	we find that 
	\be
	\log \Phi \to \log\Phi+ \ii \eta (x).
	\ee
	We shorthand $\prt_{i_1}\cdots\prt_{i_{\tm}}: = \prt^{\tm}$ so 
	\be
	\prt^{\tm} \log \Phi \to \prt^{\tm} \log\Phi+ \ii \prt^{\tm} \eta(x).
	\ee
This implies 
	\be
	P_{i_1,\cdots,i_{\tm}}(\Phi,\cdots,\prt^{\tm}\Phi) \to\re^{\ii {\tm} \eta(x)} (P_{i_1,\cdots,i_{\tm}}(\Phi,\cdots,\prt^{\tm}\Phi)
	+\ii \prt_{i_1}\cdots \prt_{i_{\tm}}\eta(x)\Phi^{\tm}).
	\ee
	Therefore it is natural to introduce the connection-like symmetric rank-m tensor gauge field $A_{i_1,\cdots,i_{\tm}}$ and higher covariant derivative 
	\bea \label{eq:gauge-covariant-D-P-A}
        D_{i_1,\cdots,i_{\tm}}[\{\Phi \}] &:=&P_{i_1,\cdots,i_{\tm}}(\Phi,\cdots,\prt^{\tm}\Phi) - \ii g A_{i_1,\cdots,i_{\tm}} \Phi^{\tm},
\eea
where we implicitly sum over all possible indices as
\bea
        \sum_{\{i_1,\cdots,i_{\tm}\}}D_{i_1,\cdots,i_{\tm}}[\{\Phi \}] &:=&
         \sum_{\{i_1,\cdots,i_{\tm}\}} \Big( P_{i_1,\cdots,i_{\tm}}(\Phi,\cdots,\prt^{\tm}\Phi) - \ii g A_{i_1,\cdots,i_{\tm}} \Phi^{\tm}\Big),
	\eea 
	which transforms covariantly under a general gauge transformation 
	\be
	\Phi \to\re^{\ii \eta (x)}\Phi.
	\ee
	This implies that the tensor gauge field $A$ transforms as 
	\bea
	A &\to& A+\frac{1}{g} \prt^{\tm} \eta.
	\eea
	More precisely, components by components, it  transforms as 
	\bea \label{eq:abelian-gauge-transformation-sym-rank-m-A}
	A_{i_1,\cdots,i_{\tm}} &\to& A_{i_1,\cdots,i_{\tm}} +\frac{1}{g} \prt_{i_1, i_2, \cdots, i_{\tm-1}, i_{\tm}} \eta := 
	A_{i_1,\cdots,i_{\tm}} +\frac{1}{g} \prt_{i_1}  \prt_{i_2}\cdots   \prt_{i_{\tm-1}} \prt_{i_{\tm}} \eta.
	\eea
So a gauge invariant term in the Lagrangian, involving the interactions between the gauge field and the scalar field, is
\be
  \left| D_{i_1,\cdots,i_{\tm}}[\{\Phi \}] \right|^2 :=
          \Big( P_{i_1,\cdots,i_{\tm}}(\Phi,\cdots,\prt^{\tm}\Phi) - \ii g A_{i_1,\cdots,i_{\tm}} \Phi^{\tm}\Big)
             \Big( P^{i_1,\cdots,i_{\tm}}(\Phi^\dagger,\cdots,\prt^{\tm}\Phi^\dagger) + \ii g A^{i_1,\cdots,i_{\tm}} (\Phi^\dagger)^{\tm}\Big) .
\ee

\subsubsection{Different meanings of ``gauging''}

We should mention that our gauging procedure is a generalization of \cite{Pretko2018jbi1807.11479}, but
our gauging procedure may be different from 
some others in the literature \cite{Vijay2016phm1603.04442, Williamson2016jiq1603.05182, Shirley2018vtc1806.08679, SeibergF2019, Radicevic2019vyb1910.06336}.
This implies that the meaning of ``\emph{gauging}'' actually is not the most unique refined statement --- there can be different ways of 
``\emph{gauging}'' although the initial global symmetry is the same.
``\emph{Gauging}'' can implies many different things:
\begin{enumerate}
\item  Promoting a global symmetry to a local symmetry, and the local symmetry fluctuation is absorbed by the gauge transformations of dynamical gauge fields.
(This is our way of gauging the higher-moment global symmetry  \cite{Wang2019aiq1909.13879, WXY3Wang2019mtt1912.13485}.)

\item Coupling the symmetry generator (the charge operator or the charge current) of the higher-moment symmetry to a background field.
And then in the partition function, one makes the background field dynamical by summing over all  the allowed background field configurations. This is also related to the orbifold procedure in field theory or string theory.

\item Gauging may also be interpreted as the condensation of the ``charged object'' $\CO$ of a global symmetry. 
The condensation means that 
the ``charged object'' becomes part of the property of the ground state wavefunction. Suppose 
 the ``charged object'' $\CO$  be a local point operator or an extended (line/surface/etc.) operator,
 then the new ground state $ | \Psi_{\text{ground state}} \rangle$ (i.e., new vacuum) with the condensed ``charged object'' $\CO$
 means that in the quantum mechanical sense,
 the ground state is in a coherent state
 \bea
\hat{\CO} | \Psi_{\text{ground state}} \rangle \propto  | \Psi_{\text{ground state}}\rangle.
 \eea
 The $\hat{\CO}$ is correspondingly a
 local or an extended quantum mechanical operator of the ``charged object'' $\CO$ in field theory. Intuitively $\hat{\CO}$ can be created and annihilated from the 
 vacuum for free -- $\hat{\CO}$ can pop out  or pop into the vacuum:
\bea
\langle \Psi_{\text{ground state}} |  \hat{\CO} | \Psi_{\text{ground state}} \rangle \propto 
\langle   \hat{\CO}  \rangle \neq 0,
\eea
which is known as the condensation in the vacuum.

\end{enumerate}
{All these above procedures are related to ``gauging,''  
although we can gauge the same initial global symmetry,
 different gauging procedures \emph{may} (or \emph{may not}) give rise to different types of gauge theories.
 In our work, we study the gauging from the perspectives of continuum field theory.
We do not yet attempt to make connections to other  ``gauging'' procedures done on the lattice \cite{Vijay2016phm1603.04442, Williamson2016jiq1603.05182, Shirley2018vtc1806.08679, Radicevic2019vyb1910.06336}, but leave for future work.}

\subsubsection{Abelian gauge field strength and tensor gauge theory}
\label{sec:AbelianF}

Follow \Sec{sec:degree-m},
to construct a rank-$({\tm}+1)$ gauge invariant abelian gauge field strength, we simply define
\bea \label{eq:abelian-gauge-field-strength-F-A}
F_{\mu,\nu,i_2,\cdots,i_{\tm}}:= \prt_{\mu} A_{\nu,i_2,\cdots,i_{\tm}} -
\prt_{\nu} A_{\mu,i_2,\cdots,i_{\tm}}.
\eea	
Here $F_{\mu,\nu,i_1,\cdots,i_{\tm}}$ is anti-symmetric respect to ${\mu \leftrightarrow \nu}$.
It is easy to check the gauge invariance of $F_{\mu,\nu,i_1,\cdots,i_{\tm}}$ under the abelian gauge transformation \Eq{eq:abelian-gauge-transformation-sym-rank-m-A}:	
\bea
F_{\mu,\nu,i_2,\cdots,i_{\tm}} \to  \prt_{\mu} (A_{\nu,i_2,\cdots,i_{\tm}} +\frac{1}{g} {\prt_{\nu} \prt_{i_2}\cdots  \prt_{i_{\tm}}}) -
\prt_{\nu} (A_{\mu,i_2,\cdots,i_{\tm}}+\frac{1}{g} {\prt_{\mu} \prt_{i_2}\cdots  \prt_{i_{\tm}}} \eta)= 
F_{\mu,\nu,i_2,\cdots,i_{\tm}} .
\eea	
It is easy to construct the gauge invariant kinetic Lagrangian term
$$
| \hat F_{\mu,\nu,i_2,\cdots,i_{\tm}}|^2 :=   F_{\mu,\nu,i_2,\cdots,i_{\tm}}   F^{{\mu,\nu,i_2,\cdots,i_{\tm}}}.
$$

\subsubsection{Non-abelian gauge field strength and tensor gauge theory}
\label{sec:NAbelianF}

We can also promote the abelian gauge field strength \Eq{eq:abelian-gauge-field-strength-F-A} to a non-abelian gauge field strength,
follow the trick of \Refe{Wang2019aiq1909.13879} by gauging the 
ordinary 0-form $\Z_2^C$-charge conjugation global symmetry. The $\Z_2^C$ acts on the rank-m tensor gauge field via: 
\bea \label{eq:Z2C-on-sym-rank-m-A}
	A_{i_1,\cdots,i_{\tm}} \to - A_{i_1,\cdots,i_{\tm}}. 
\eea
By promoting the global $\Z_2^C$ to a local symmetry,
we introduce a new 1-form $\Z_2^C$-gauge field $C$ coupling to the
0-form symmetry $\Z_2^C$-charged object $A_{i_1,\cdots,i_{\tm}}$ with a new $g_c$ coupling.
The $\Z_2^C$ local gauge transformation is: 
\bea \label{eq:Z2C-vector-gauge}
A_{i_1,\cdots,i_{\tm}}  \to  \re^{\ii \gamma_c(x)} A_{i_1,\cdots,i_{\tm}}, \quad 
C_\nu \to C_\nu +\frac{1}{g_c} \prt_\nu \gamma_c(x).
\eea
Note $A_{i_1,\cdots,i_{\tm}}$ is real-valued, so a generic $\re^{\ii \gamma_c(x)}$ complexifies the $A_{i_1,\cdots,i_{\tm}}$. 
However, what we can do is restricting  gauge transformation so it is only $\Z_2^C$-gauged (not $\U(1)^C$-gauged)
 \bea \label{eq:Z2-phase}
 \re^{\ii \gamma_c(x)} := (-1)^{\gamma_c'(x)} \in \{\pm 1\},
 \eea
 so ${\gamma_c'(x)}$ is an integer and $A_{i_1,\cdots,i_{\tm}}$ stays in real. 
 Thus ${\gamma_c'(x)}$ jumps between even or odd integers in $\Z$, while the 
 $\Z_2^C$-gauge transformation can be suitably formulated on a lattice. 
 We can directly rewrite the above \Eq{eq:Z2C-vector-gauge} on a simplicial complex or a triangulable spacetime manifold.
Follow \Refe{Wang2019aiq1909.13879}, we also define a new covariant derivative with respect to $\Z_2^C$:
\bea \label{eq}
D_\mu^c :=(\prt_\mu - \ii g_c C_\mu).
\eea
We need to combine $ \U(1)_{x_{(n)}}$-gauge transformation \Eq{eq:abelian-gauge-transformation-sym-rank-m-A} and $\Z_2^C$-gauge transformation \Eq{eq:Z2C-vector-gauge} to:
 \bea \label{eq:vector-gauge-charge}
  A_{i_1,\cdots,i_{\tm}}  &\to&  \re^{\ii \gamma_c(x)} A_{i_1,\cdots,i_{\tm}}+ \frac{1}{({\tm}!) g}(D_{(i_1}^c D_{i_2}^c \dots D_{i_{\tm})}^c) ( \eta_v(x)),
   \nn\\
    C_\nu &\to& C_\nu +\frac{1}{g_c} \prt_\nu \gamma_c(x).
\eea

Here $(D_{(i_1}^c D_{i_2}^c \dots D_{i_{\tm})}^c):=
(D_{i_1}^c D_{i_2}^c \dots D_{i_{\tm}}^c
+
D_{i_2}^c D_{i_1}^c \dots D_{i_{\tm}}^c
+\dots)$ contains the permutation $({\tm}!)$-terms, which means to be a symmetrization over the subindices under the lower bracket  $({i_1,\cdots,i_{\tm}})$.

We thus can promote the abelian gauge field strength \Eq{eq:abelian-gauge-field-strength-F-A}'s $F_{\mu,\nu,i_2,\cdots,i_{\tm}}$ into a new non-abelian gauge field strength 
$\hat F^c_{\mu,\nu,i_2,\cdots,i_{\tm}}$
after gauging  $\Z_2^C$: 
\bea
\boxed{\hat F^c_{\mu,\nu,i_2,\cdots,i_{\tm}}  :=D_\mu^c A_{\nu,i_2,\cdots,i_{\tm}}  -D_{\nu }^c A_{\mu,i_2,\cdots,i_{\tm}}
:=(\prt_\mu - \ii g_c C_\mu) A_{\nu,i_2,\cdots,i_{\tm}} -(\prt_{\nu} - \ii g_c C_\nu ) A_{\mu,i_2,\cdots,i_{\tm}} }. 
\eea
This $\hat F^c_{\mu,\nu,i_2,\cdots,i_{\tm}}$ is covariant under the gauge transformation \Eq{eq:vector-gauge-charge}:
$\hat F^c_{\mu,\nu,i_2,\cdots,i_{\tm}}\to \re^{\ii \gamma_c(x)}  \hat F^c_{\mu,\nu,i_2,\cdots,i_{\tm}}$.
It is obvious that we can construct the gauge invariant kinetic Lagrangian term
$$
| \hat F^c_{\mu,\nu,i_2,\cdots,i_{\tm}}|^2 :=  \hat F^c_{\mu,\nu,i_2,\cdots,i_{\tm}}  \hat F^{\dagger c\; {\mu,\nu,i_2,\cdots,i_{\tm}}},
$$
pairing $\hat F^c$ with its complex conjugation $ \hat F^{\dagger c}$.
We can propose a schematic path integral form:
\bea \label{eq:U(1)poly-Z2C-gauge}
\bZ_{\underset{\text{asym-BF}}{\text{rk-m-sym-$A$}}}: =
\int[\cD A_{i_1,\cdots,i_{\tm}}][\cD B][\cD C]\exp(\ii \int_{M^{d+1}}  \Big( 
|\hat F^c_{\mu,\nu,i_2,\cdots,i_{\tm}}|^2 \dd^{d+1}x  + \frac{ 2}{2 \pi} B \dd C\Big)) \cdot \omega_{d+1}(\{C_I\}).
\eea
Based on the knowledge of the lower degree-1 polynomial gauge structure \Eq{eq:Z2CU1xn-gauge} and the higher-moment degree-(m$-1$) polynomial symmetry \Eq{eq:U1-polynomial-product-group}, 
we can denote the gauge structure for \Eq{eq:U(1)poly-Z2C-gauge} as
\bea \label{eq:Z2CU1pxn-gauge}
\left[\Z_2^C \ltimes \Big(     \U(1)_{x_{ n \choose {\tm-1}}^{{\tm-1}}} \Big) \right]. 
\eea
If only the rank-m symmetry tensor gauge field $A_{i_1,\cdots,i_{\tm}}$ is kept, then the lower degree polynomial symmetry in \Eq{eq:U1-polynomial-product-group}, say
$\U(1)_{x_{n \choose M}^{M}}$ for $0\leq M \leq \tm-2$ would be \emph{neither} a survived global symmetry \emph{nor} a gauged symmetry.

More generally, we introduce the index $I$ for specifying the different copies/layers of tensor gauge theories,
\begin{multline} \label{eq:U(1)poly-Z2C-gauge}
\bZ_{\underset{\text{asym-BF}}{\text{rk-m-sym-$A$}}}: =
\int
(\prod_{I=1}^{N}[\cD A_{I,{i_1,\cdots,i_{\tm}}}] [\cD B_I]  [\cD C_I])\\
\exp(\ii \int_{M^{d+1}} \dd^{d+1}x \Big( 
\sum_{I=1}^N  |\hat F^c_{I,\mu,\nu,i_2,\cdots,i_{\tm}}|^2  + \frac{ 2}{2 \pi}  \sum_{I=1}^{N} B_I \dd C_I\Big)) \cdot \omega_{d+1}(\{C_I\}).
\end{multline} 
where the level-2 BF theory is used to constrain the flat $C$ gauge field to be a $\Z_2$-valued 1-form gauge field via
a $\Z_2$-valued $(d-1)$-form gauge field $B$, 
based on the trick of \cite{Wang2019aiq1909.13879}.
The cocycle
$\omega_{d+1} \in \rH^{d+1}((\Z_2^C)^N, \R/\Z)$ is a group cohomology data \cite{DijkgraafWitten1989pz1990} that we apply its continuum field theory formulation   \cite{1602.05951WWY, Putrov2016qdo1612.09298, Wang2018edf1801.05416, Wang1901.11537WWY}  (see the overview \cite{Wang2019aiq1909.13879}).
The cocycle $\omega_{d+1}$ couples different copies/layers of tensor gauge theories together, which can be viewed as interlayer interaction effects.  

Above we formulate a general degree polynomial as a higher moment global symmetry and construct the field strength $F$ for the gauge theory. 
	Our theory presented above in \Sec{sec:degree-m} is general. 
	Let us take two special examples in the next subsections, for polynomial of degree 1 in \Sec{sec:degree1} and degree 2 in \Sec{sec:degree2}.

\subsection{Polynomial with degree 1: Vector symmetry}
\label{sec:degree1}
For a vector global symmetry of a polynomial with degree 1, the symmetry transformation on the scalar field and
the invariant quantity under this transformation are as follows:
\bea
\label{eq:degree-1-sym}
\Phi&\to&\re^{\ii Q(x)}\Phi =   \re^{\ii (\Lambda_{i}  x_i + \Lambda_0)}\Phi,\\
	\log \Phi &\to& \log\Phi+ \ii Q(x) = \log\Phi+ {\ii (\Lambda_{i}  x_i + \Lambda_0)},\\
		\prt_{x_i}\prt_{x_j}\log\Phi &=&
		\frac{P_{x_i,x_j}(\Phi,\prt\Phi,\prt^2\Phi)}{\Phi^2}= \frac{{\Phi \prt_{x_i} \prt_{x_j} \Phi -( \prt_{x_i} \Phi)( \prt_{x_j} \Phi) }}{\Phi^2} \to \prt_{x_i}\prt_{x_j}\log\Phi ,\\
				\prt_{i}\prt_{j}\log\Phi &=&\frac{{\Phi \prt_{i} \prt_{j} \Phi -( \prt_{i} \Phi)( \prt_{j} \Phi) }}{\Phi^2}.
	\eea 
	In the last line, we simply shorthand $x_i, x_j$ as $i,j$. 
To gauge,	we rewrite $Q(x)$ as a local gauge parameter $\eta(x)$,
\bea
\prt_{i}\prt_{j}\log\Phi &\to& \prt_{i}\prt_{j}\log\Phi  +  \ii \prt_{i}\prt_{j}\eta(x).
\eea
This implies that we can write the gauge covariant operator $D_{i,j}[\{\Phi \}]$ via:
	\bea
	P_{i,j}(\Phi,\prt\Phi,\prt^2\Phi) &:=& ({{\Phi \prt_{i} \prt_{j} \Phi -( \prt_{i} \Phi)( \prt_{j} \Phi) }}) \to\re^{\ii 2 \eta(x)} (P_{i,j}(\Phi,\prt\Phi,\prt^2\Phi)
	+\ii \prt_{i} \prt_{j}\eta(x)).\\
        	A_{i,j}  &\to &A_{i,j} +\frac{1}{g} \prt_{i} \prt_{j} \eta.\\
	        D_{i,j}[\{\Phi \}] &:=&P_{i ,j}(\Phi,\prt\Phi,\prt^2\Phi) - \ii g A_{i,j} \Phi^2 = ({{\Phi \prt_{i} \prt_{j} \Phi -( \prt_{i} \Phi)( \prt_{j} \Phi) }} - \ii g A_{i,j} \Phi^2).
	\eea
Physically we may define the symmetry transformation
\bea
\re^{\ii \Lambda\cdot x}=	\re^{\ii 2 \pi (\lambda^{-1})\cdot x}.
\eea
So the $\lambda$ is a vector of an effective wavelength while 
$\vec \Lambda = |\Lambda| \hat \Lambda  = \frac{2 \pi }{|\lambda|} \hat{\lambda}$, with the unit vector $\hat \Lambda= \hat{\lambda}$.
So a gauge invariant term can be
$\left| D_{i,j}[\{\Phi \}] \right|^2$.

See another route of pursuit for the vector global symmetry recently by Seiberg \cite{SeibergF2019}.

\subsection{Polynomial with degree 2: Higher-moment symmetry}
\label{sec:degree2}

For a vector global symmetry of a polynomial with degree 2, the symmetry transformation on the scalar field and
the invariant quantity under this transformation are as follows:
\bea
\label{eq:degree-2-sym}
\Phi&\to&\re^{\ii Q(x)}\Phi =   \re^{\ii ( \Lambda_{i,j}  x_i x_j  + \Lambda_{i}  x_i + \Lambda_0) }\Phi,\\
	\log \Phi &\to& \log\Phi+ \ii Q(x) = \log\Phi+ {\ii ( \Lambda_{i,j}  x_i x_j  + \Lambda_{i}  x_i + \Lambda_0)},\\
		\prt_{i}\prt_{j}\prt_{k}\log\Phi &=&
		\frac{P_{i,j,k}(\Phi,\cdots,\prt^3\Phi)}{\Phi^3}\\
		&:=& \frac{{\Phi^2 (\prt_{i} \prt_{j} \prt_{k} \Phi )
		-\Phi\big( (\prt_{k} \Phi)( \prt_{i} \prt_{j}  \Phi )+  (\prt_{i} \Phi)( \prt_{j} \prt_{k}  \Phi ) +  (\prt_{j} \Phi)( \prt_{i} \prt_{k}  \Phi )\big)
		  } +  2({ ( \prt_{i} \Phi)( \prt_{j} \Phi)(\prt_{k} \Phi ) }) }{\Phi^3}. \nn\\
		  &:=& 
		  \frac{{{\Phi^2 (\prt_{i} \prt_{j} \prt_{k} \Phi )
		-3\Phi\big( \prt_{(k} \Phi \prt_{i} \prt_{j)}  \Phi \big)
		  } +  2{ ( \prt_{i} \Phi)( \prt_{j} \Phi)(\prt_{k} \Phi ) } }}{\Phi^3}. \nn
			\eea 
This implies that we can write the gauge covariant operator $D_{i,j,k}[\{\Phi \}]$ via:
	\bea
	P_{i,j,k}(\Phi,\cdots,\prt^3\Phi) &:=& {{\Phi^2 (\prt_{i} \prt_{j} \prt_{k} \Phi )
		-\Phi\big( (\prt_{k} \Phi)( \prt_{i} \prt_{j}  \Phi )+  (\prt_{i} \Phi)( \prt_{j} \prt_{k}  \Phi ) +  (\prt_{j} \Phi)( \prt_{i} \prt_{k}  \Phi )\big)
		  } +  2{ ( \prt_{i} \Phi)( \prt_{j} \Phi)(\prt_{k} \Phi ) } }\nn\\
		   &:=&
		   {{\Phi^2 (\prt_{i} \prt_{j} \prt_{k} \Phi )
		-3\Phi\big( \prt_{(k} \Phi \prt_{i} \prt_{j)}  \Phi \big)
		  } +  2{ ( \prt_{i} \Phi)( \prt_{j} \Phi)(\prt_{k} \Phi ) } }\nn\\
		   &\to&\re^{\ii 3 \eta(x)} (P_{i,j,k}(\Phi,\cdots,\prt^3\Phi)
	+\ii \prt_{i} \prt_{j}\prt_{k}\eta(x)).\\
        	A_{i,j,k}  &\to &A_{i,j,k} +\frac{1}{g} \prt_{i} \prt_{j}  \prt_{k} \eta.\\
	        D_{i,j,k}[\{\Phi \}] &:=&P_{i ,j,k}(\Phi,\cdots,\prt^3\Phi) - \ii g A_{i,j,k} \Phi^3.
	\eea	
The above we use the symmetrized tensor notation:
$T_{(i_1i_2\cdots i_k)} = \frac{1}{k!}\sum_{\sigma\in \mathfrak{S}_k} T_{i_{\sigma 1}i_{\sigma 2}\cdots i_{\sigma k}}$,
with parentheses $(ijk)$ around the indices being symmetrized.
The $\mathfrak{S}_k$ is the symmetric group of $k$ symbols.
So a gauge invariant term can be
$\left| D_{i,j,k}[\{\Phi \}]\right|^2$.

\section{Vector Charge, Tensor Charge and General Higher-Moment Symmetry}

\subsection{Vector charge}
\label{sec:Vectorcharge}

We may also consider more general higher moment conservation laws with a set of number $r$ fields $\Phi_1,\cdots \Phi_r$ and gauge transformations
	\be
	\Phi_I \to \re^{\ii Q_I(x)}\Phi_I
	\ee
	where 
	\be
	Q_I\in V_I \subset \oplus_{I=1}^r V_I = V\subset \mathbb{R}^r\otimes \mathbb{R}[x_1,\cdots,x_n]= \oplus_{I=1}^r \mathbb{R}[x_1,\cdots,x_n].
	\ee
	The $V_I$ denotes the vector space where the polynomial $Q_I=Q_I(x)$ lives in. A special case is 
	$V_I =\mathbb{R}[x_1,\cdots,x_n].$ For this vector space $\mathbb{R}[x_1,\cdots,x_n]$, 
	we have the vector addition in terms of the polynomial addition, while 
	we have the scalar multiplication in terms of the scalar in the real number $\mathbb{R}$ multiplying by the polynomial.
	
	Note that the full vector space $V$ is fully characterized by another vector space $D$ of differential operators, which annihilate the space $V$. 
	This space $D_I$ is not finite dimensional, but we may take a finite dimensional subspace $\tilde{D}$ generating the vector space $D$. Namely, we can take differential operators 
	$\fD^I_J$ 
	such that 
	\bea
	\sum_{I}\fD^I_J Q_I=0
	\eea 
	for any $Q_I\in V$. If 
	\bea
	\sum_{I}\fD^I Q_I=0
	\eea for any $Q$ we have 
	\bea
	\fD^I=\sum a^J\fD^I_J
	\eea for some $a^J$. 
	In the previous example, we took homogeneous polynomials of degree $m$.

	Following the same logic, we may show that each elements in $\tilde{D}$ gives an invariant field strength. 
	To gauge this symmetry, we need to introduce gauge fields which one to one corresponds to elements in $\tilde{D}$.
	The gauge transformation law is transparent: 
	they are just differential operators $\fD^I_J $ in $\tilde{D}$ acting on the gauge variational parameter say $\eta_I$. 
	
	Let us be more concrete: under the gauge transformation $\Phi_I \to \re^{\ii Q_I(x)}\Phi_I$,	
	we have 
	$
	\log\Phi_I \to \log\Phi_I +\ii Q_I(x)
	$
	and
	\bea
	\sum_{I}\fD^I_J\log\Phi_I \to \fD_J\log\Phi_I
	\eea
	as $\sum_{I}\fD^I_JQ_I=0$.
	We can compute 
	\bea
	\sum_{I}\fD^I_J\log\Phi_I = \frac{P_J}{\CQ_J}
	\eea
	where $P_J$ is a polynomial in fields $\Phi_I$ and their derivatives, and $\CQ_J$ is a monomial in $\Phi_I$. 
	Namely,  $\CQ_J =\prod_{I} (\Phi_I)^{(n^I_J)}$ is a product of  $\Phi_I$, where $n^I_J$ is an integer power of some $\Phi_I$.
	
Therefore we can see that because the denominator $\CQ_J$ as a monomial transforms in the following way, same for the 
	numerator $P_J$:
	\bea
        \CQ_J &\to& e^{\ii \zeta_J(x)}\CQ_J,\\
	P_J &\to& e^{\ii \zeta_J(x)}P_J,
	\eea
	for some polynomial $ \zeta_J= \zeta_J(x)$ depending on our data.  
Each of these $P_J$ corresponds to a covariant derivative under a general gauge transformation 
\bea
	\Phi_I\to e^{\ii \eta_I(x)} \Phi_I.
\eea
We can compute 
\bea	
	\sum_{I}\fD^I_J\log\Phi_I = \frac{P_J}{\CQ_J}+	\ii\sum_{I}\fD^I_J\eta_I,
\eea
hence $P_J$ transform as
\bea
	P_J\to e^{\ii\zeta_J(x)}P_J +\ii\sum_{I}\fD^I_J\eta_I \CQ_J.
\eea
	Therefore we introduce a new covariant derivative 
	\bea
	D_J[\{\Phi_I \}]  :=R_J \equiv P_J - \ii g A_J \CQ_J
	\eea
where under the gauge transformation of $\Phi_I\to e^{\ii \eta_I(x)} \Phi_I$,
we have the gauge transformation of: 
	\bea
	A_J \to A_J+\frac{1}{g}\sum_{I}\fD^I_J\eta_I.
	\eea 
Then $R_J$ transforms as $R_J\to e^{\ii \zeta_J (x)}R_J$. We construct the gauge invariant matter-gauge field interaction term 
in the Lagrangian
\bea
|D_J[\{\Phi_I \}]|^2  := |R_J|^2 :=  (R_J) (R_J^{\dagger}) =( P_J - \ii g A_J \CQ_J) ( P_J^{\dagger} + \ii g A_J \CQ_J^{\dagger}).
\eea

\subsection{Tensor charge}
\label{sec:Tensorcharge}

It is embarrassingly easy to generalize to	this tensor-index complex scalar field for the global symmetry transformation:
	\bea
	\Phi_{I_1, \dots, I_M}\to \re^{\ii Q_{I_1, \dots, I_M}(x)} \Phi_{I_1, \dots, I_M},
	\eea
	and the gauge symmetry transformation with a local dependent gauge parameter ${\eta_{I_1, \dots, I_M}(x)}$:
	\bea
	\Phi_{I_1, \dots, I_M}\to \re^{\ii \eta_{I_1, \dots, I_M}(x)} \Phi_{I_1, \dots, I_M}.
	\eea
Nonetheless, we just need to give a one-to-one map $(I_1, \dots, I_M) \to {I}$,
so the tensor indices can be mapped to a vector index, which transforms the above two equations to:
$$
\Phi_{I}\to e^{\ii \alpha_{I}(x)} \Phi_{I}
$$
and
$$
\Phi_{I}\to e^{\ii \eta_{I}(x)} \Phi_{I}
$$
respectively. Thus the tensor charge higher-moment global symmetry can be treated as the same way
as  the vector charge higher-moment global symmetry  in \Sec{sec:Tensorcharge}
under the one-to-one map $(I_1, \dots, I_M) \to {I}$.

\subsection{Example 1: Vector charge with an exclusive degree-1 polynomial}
	\label{sec:ex1}
	
	As a special case, we can recover the vector charged tensor gauge theory by taking 
	the vector space of degree-1 polynomial
	\bea
	V_I= \{1,x_{I-1},x_{I+1} \} 
	\eea
  which means that it is spanned by the vectors of $1$, $x_{I-1}$, and $x_{I+1}$.
The global symmetry acts as
\be \label{eq:Example-1-vector-global}
\Phi_I(x) \to \re^{\ii Q_I(x)}\Phi_I(x)  =  \re^{\ii Q_I(x_{I-1},x_{I+1})}\Phi_I(x)
 =  \re^{\ii (\Lambda_{I+1} x_{I-1} - \Lambda_{I-1} x_{I+1}+ \Lambda_0)}\Phi_I(x).
\ee
  Here we may define 
  $$x_{I+l'}: = x_{{I+l'} \mod n}, \text{with the subindex where } {I+l'} : ={I+l' \mod n}.$$ 
In fact, our specific example here is a generalization of one example in Pretko's \cite{Pretko2018jbi1807.11479}.\footnote{Our result 
in \Eq{eq:Example-1-vector-global} generalizes Pretko's
\be 
\Phi_I(x) \to \re^{\ii \sum_{J,K}\varepsilon_{IJK} \Lambda_J x_K }\Phi_I(x),
\ee
where $\varepsilon_{IJK}=\varepsilon^{IJK}$ is just a Levi-Civita symbol, or a so-called alternating tensor.
}
We may call this type of $Q_I(x)= Q_I(x_{I-1},x_{I+1})$ as an exclusive polynomial
which the $Q_I$ excludes the $x_I$ dependency, thus it is $x_I$ independent.

   The vector space $V$ is fully characterized by another vector space $D$ of differential operators, which annihilate $V$ by differential. 
This space $D_I$ is not finite dimensional, but we may take a finite dimensional subspace $\tilde{D}$ generating the vector space $D$,
	here\footnote{In this notation below
	$\Phi_I \to \re^{\ii Q_I(x)}\Phi_I  =  \re^{\ii Q_I(x_{I-1},x_{I+1})}\Phi_I$, we just focus on the dependence of $x$ only on $Q_I(x)$, not $\Phi_I$.}
 \bea
 \tilde{D}= \{\prt_{(i=I)} \Phi_I, \Phi_J \prt_{(j=J)} \Phi_K+\Phi_K\prt_{(k=K)} \Phi_J  \},
 \eea 
 where $j=J$ and $k=K$ are related by 
 $$
 j=k \pm 1 \mod n.
 $$
 Here the spacetime index ($i,j,k,\dots$) and the internal vector index ($I,J,K,\dots$) of $\Phi$ fields are locked.

Furthermore, we can effectively construct the gauge theory explicitly, given by the rule of gauge principle.
For this special case, we can recover the vector charged tensor gauge theory 
and covariant derivatives:
\bea
 \{ \prt_{(i=I)} \Phi_I - \ii g A_I\Phi_I, \quad \Phi_J \prt_{(j=J)}\Phi_K+\Phi_K\prt_{(k=K)}\Phi_J - \ii  g A_{(j=J)(k=K)}\Phi_J\Phi_K \}.
\eea 
In short, by locking $i=I$, $j=J$ and $k=K$, we simply write
\bea
 \{ \prt_{i} \Phi_i - \ii g A_i\Phi_i, \quad \Phi_j \prt_{j}\Phi_k+\Phi_k\prt_{k}\Phi_j - \ii  g A_{j k}\Phi_J\Phi_K \}.
\eea 	
	We can effectively construct everything explicitly, given the rule of gauge transformations:
	\bea
	\Phi_{I} &\to& e^{\ii \eta_{I}(x)} \Phi_{I},\\
	A_{j k} &\to& A_{j k} +\frac{1}{g}( \prt_{j} \eta_k + \prt_{k} \eta_j).
	\eea


\subsection{Example 2: Vector charge with an inclusive degree-1 polynomial}
	\label{sec:ex2}

Let us consider another simple example: Given fields $\Phi_1$, $\Phi_2$, $\dots$,
and the higher-moment global symmetry: 
\bea\Phi_i\to e^{\mathrm{i}\Lambda \; x_i}\Phi_i.
\eea
We may call this type of $Q_I(x)= Q_I(x_{I})$ as an inclusive polynomial
which the $Q_I$ include only the $x_I$ dependency.
We have an invariant Lagrangian term $|\Phi_2\partial_1\Phi_1-\Phi_1\partial_2 \Phi_2|^2$.
We can introduce a tensor connection field $A_{12}$, then the covariant derivative type of Lagrangian term 
$
|\Phi_2\partial_1\Phi_1-\Phi_1\partial_2\Phi_2-\ii g A_{12}\Phi_1\Phi_2|^2.
$
More generally, we have
\bea
|\Phi_i \partial_j \Phi_j-\Phi_j\partial_i \Phi_i-\ii g A_{ij}\Phi_i\Phi_j|^2
\eea
invariant under a general gauge transformation
\bea
\Phi_j  &\to& e^{\mathrm{i}\eta_j(x)}\Phi_j,\\
A_{ij} &\to& A_{ij}+\frac{1}{g}(\partial_i\eta_i-\partial_j\eta_j),
\eea
where $i,j$ can be any coordinate since we have this specific global symmetry: $\Phi_i\to e^{\mathrm{i}\Lambda \; x_i}\Phi_i$ for any $x_i$.
Importantly, the generic gauge field $A_{ij}$ is \emph{not} symmetric 
under $i\leftrightarrow j$.\footnote{$A_{ij}$  can be made symmetric if we revise the transformation law, 
for a specific pair of $(i,j)$,
such that $\Phi_i\to \re^{\mathrm{i}\Lambda \; x_i}\Phi_i$ and
$\Phi_j\to \re^{-\mathrm{i}\Lambda \; x_j}\Phi_j$, so that the Lagrangian term
$$
|\Phi_i \partial_j \Phi_j+\Phi_j\partial_i \Phi_i-\ii g A_{ij}\Phi_i\Phi_j|^2
$$
invariant under a general gauge transformation
$$
\Phi_j  \to \re^{\mathrm{i}\eta_j(x)}\Phi_j,
\quad
A_{ij} \to A_{ij}+\frac{1}{g}(\partial_i\eta_i+\partial_j\eta_j).
$$}
This example reveals that the generic higher-moment global symmetry for a vector-index charge field, after gauging, does not yield a symmetric tensor gauge field
.

\subsection{Example 3: Vector charge with a mixed degree-1 polynomial}
	\label{sec:ex3}
	
Consider the vector-index charge fields:  $\Phi_j$.
Consider the higher-moment global symmetry: 
$$
\Phi_j\to {\re}^{\mathrm{i}\Lambda x_1} \Phi_j,
$$
where $j=1,2,\dots$
We have an invariant Lagrangian term $|\Phi_2\partial_1 \Phi_1-\Phi_1\partial_1 \Phi_2|^2$,
and other invariant Lagrangian terms $|\Phi_j \partial_1 \Phi_i-\Phi_i \partial_1 \Phi_j|^2$.
%
We can introduce a tensor gauge connection field $A_{ij}$, then the covariant derivative 
\bea
\Phi_j\partial_1\Phi_i-\Phi_i\partial_1\Phi_j-iA_{ij}\Phi_i\Phi_j
\eea
is invariant under a general gauge transformation
\bea
\Phi_j &\to& \re^{\mathrm{i}\eta_j(x)}\Phi_j,\\
A_{ij} &\to & A_{ij}+(\partial_1\eta_i-\partial_1\eta_j).
\eea
Again $i,j$ can be any coordinate since we have this specific global symmetry: $\Phi_i\to \re^{\mathrm{i}\Lambda \; x_1}\Phi_i$ for any $x_i$.
Importantly, similar to \Sec{sec:ex3}, the generic gauge field $A_{ij}$ is \emph{not} symmetric 
under $i\leftrightarrow j$

\section{Generalizing Noether's Theorem
for {Higher-Moment Global Symmetry}}

Suppose we have a set of $r$ fields 
$\Phi_I \,(1\leq I\leq r)$, a Lagrangian term $L$ which is invariant under a global transformations
$
\Phi_I \to \re^{\mathrm{i} Q_I(x)}\Phi_I
$
where $(Q_1,\cdots,Q_r)\in V\subset \oplus _{I=1}^r \mathbb{R}[x_1,\cdots,x_d]$ is a specified vector space of allowed polynomials.

Noether's theorem guarantees that we have a conserved current corresponding to each global symmetry. Suppose the constant U(1) transformation for each field is a global symmetry, Noether's theorem says that we have a one form current
\bea
j_I=j_{I\mu} \dd x^\mu,
\eea such that under the general infinitesimal variation 
\bea
\Phi_I\to \re^{\mathrm{i}\epsilon\alpha_I (x)} \Phi_I,
\eea
the Lagrangian density transforms as
\bea
\delta\mathcal{L}=\epsilon \alpha_I \wedge \star j_I.
\eea 
The $\epsilon$ is an infinitesimal variational parameter.
Here $\rho_I=j_{I0}$ is the spatial density of the conserved charge. 

Now let us take $\alpha_I=Q_I(x)$ as the higher-moment global symmetry polynomial for 
$$
\Phi_I \to \re^{\mathrm{i} Q_I(x)}\Phi_I.
$$
As we said earlier, we see that $\int \delta \mathcal{L}$=0, and therefore 
\bea
\int _{\text{space}}\sum_I \rho_I Q_I=\int _{\text{space}}\sum_I j_{I0} Q_I
\eea
is a conserved charge.
That is, we have a conserved charge for each of the global symmetry we have, their number is precisely the dimension of the vector space $V$ we started with.

By doing the above calculation, we need to be careful about the boundary conditions of the space manifold,
or the infinite faraway field configurations of the space manifold. In most cases, we can assume that the density
of field configurations decays sharply at the infinite faraway.

Let us take $Q_I(x)$ is a polynomial over the spatial coordinates. 
Here are some examples:\\[-9mm]
\begin{enumerate}[
label=\textcolor{blue}{(\arabic*)}., ref=\textcolor{blue}{(\arabic*)}]
\item For a single field $\Phi$, when $Q(x)=\Lambda$ is a constant, we have the usual Noether's theorem for the ordinary U(1) global symmetry,
with a conserved charge:
\bea
\int _{\text{space}} \rho =\int _{\text{space}} j_{0} .
\eea
\item For a single field $\Phi$, when $Q(x)={ \Lambda_{i}  x_i + \Lambda_0} $ is a linear degree-1 polynomial, we have a conservation 
theorem for the vector U(1) global symmetry.
This coincides an example of Pretko's \cite{Pretko2016lgv1606.08857}.
\bea
&&\Lambda_0\text{'s}: \int _{\text{space}} \rho =\int _{\text{space}} j_{0} .\\
&&\Lambda_i\text{'s}:  \int _{\text{space}} \rho   x_i
=\int _{\text{space}} j_{0} x_i.
\eea
There are the same number of conserved charges as the dimensions of the vector space 
(the independent parameters $\Lambda_0$ and $\Lambda_i$ of the degree-1 polynomials).
\item 
For a single field $\Phi$, when 
we follow \Eq{eq:Qx-general} with
	$$
Q(x) :=	(\Lambda_{i_1,\dots,i_{\tm-1}}  x_{i_1} \dots x_{i_{\tm-1}} +\dots+ \Lambda_{i,j}  x_ix_j + \Lambda_{i}  x_i + \Lambda_0\big)
	$$
of a degree-$({\tm-1})$ polynomial, we have a conservation theorem for all independent 
$\Lambda_{i_1,\dots,i_{k}}$
\bea
&&\Lambda_{i_1,\dots,i_{k}} \text{'s}:  \int _{\text{space}} (\rho) \cdot ( x_1 \dots x_k)
=\int _{\text{space}} (j_{0})  \cdot ( x_1 \dots x_k).
\eea
for the higher-moment U(1) global symmetry.
There are the same number of conserved charges as the dimensions of the vector space 
(the independent parameters $\Lambda_{i_1,\dots,i_{k}}$).
\item 
For a vector-index field $\Phi_I$,  
with
	$$
Q_I(x) :=	(\Lambda_{I; i_1,\dots,i_{\tm-1}}  x_{i_1} \dots x_{i_{\tm-1}} +\dots+ \Lambda_{ I;  i,j}  x_ix_j + \Lambda_{I;  i}  x_i + \Lambda_{I;  0}\big),
	$$
There are the same number of conserved charges as the dimensions of the vector space 
(the independent parameters $\Lambda_{I; i_1,\dots,i_{k}}$).	
\item 
For a tensor-index field $\Phi_{I_1, \dots, I_M}$,  we can map to a vector-index field
$\Phi_I$ by a one-to-one map $(I_1, \dots, I_M)$ $\to$ ${I}$, thus the result follows from the previous remark.
\end{enumerate}
In all cases, if we have additional constraints 
(such as from the constraint of field strength, 
say the electric tensor in  \cite{Pretko2016lgv1606.08857} to be traceless,
say $\tilde {\rm E}_{j}^j=0$ for $\tilde {\rm E}_{ij} = -\partial_0A_{ij}+\partial_i\partial_j A_0 = -\partial_tA_{ij}+\partial_i\partial_j A_0$ for \cite{Wang2019aiq1909.13879}'s notation),
then we have additional new conservation laws, not accounted by the previously counted number of conserved laws as the dimensions of the vector space of $Q_I(x)$.
 


\section{Conclusion, and Relations to Algebraic Geometry}

In this section, we bridge the relations between our theories (both the matter or the gauge theories) by physics construction and 
the algebraic geometry in mathematics.
We conclude with some final comments. 

\subsection{Algebraic (affine) Variety and Subvariety}
\label{eq:AlgebraicVariety}

In mathematics, the polynomials are related to  geometric objects called the algebraic variety. More precisely, (affine) varieties are defined 
as the solutions of polynomial equations. The morphisms between them are maps defined by polynomials. 
Here we review their basic definitions for both physicists and mathematicians:
	\begin{defn}
		An affine algebraic variety over real numbers $\R$ is the zero-locus in the affine space $\R^n$ of some finite family of polynomials of n variables with coefficients in $\R$.
	\end{defn}

\begin{defn}
	A morphism, or a regular map, of affine varieties is a function between affine varieties which is polynomial in each coordinate: more precisely, for affine varieties $V\subseteq \R^n$ and $W\subseteq \R^m$, a morphism from V to W is a map $\phi : V \longrightarrow W$ of the form $\phi(a_1, ..., a_n) = (f_1(a_1, ..., a_n), ..., f_m(a_1, ..., a_n))$, where $f_i \in \R[X_1, ..., X_n]$ for each $i = 1, ..., m$.
	Here $(a_1, ..., a_n) \in V\subseteq \R^n$
	and $\phi(a_1, ..., a_n) \in W\subseteq \R^m$.
	
\end{defn}

\begin{defn}
	Given two affine varities $V,W\subseteq \R^n$, V is called a subvariety of W, if $V\subseteq W$ as subsets of $\R^n$.
\end{defn}

\begin{defn}
	Two affine varieties V and W are isomorphic if there exist morphisms $\phi:V\longrightarrow W$ and $\psi:W\longrightarrow V$ such that $\psi\circ\phi=\rm{id}_V$ and $\phi\circ\psi=\rm{id}_W$, where $\rm{id}$ is the identity map.
\end{defn}
	For example,  $x^2+y^2=1$ defines the unit circle and 
	$
	\frac{t^2}{4}+w^2=1
	$
	defines an ellipse, both on the plane $ \R^2$. 
	There is a polynomial map 
	$
	t\to 2x \,, w\to y
	$
	which identifies circle and ellipse, with inverse given by a rescaling
	$
	x\to t/2 \,, y\to w
	$ therefore in algebraic geometry they are isomorphic.
	
	In our setting, the higher-moment global symmetry transformations are given by polynomials on the space (here we focus on the Cartesian
	$ \R^n$ or $\R^d$ stated since \Sec{sec:intro}), 
and the contours (or constant hypersurfaces) are given by solutions of polynomials: they are subvarieties of our space (here on the Cartesian
	$ \R^n$ or $\R^d$). 
	
	We should mention that \Refe{Vijay2015mka1505.02576VijayHaahFu, Vijay2016phm1603.04442} has a different look on the 
	algebraic variety:  the topological degeneracy of the gapped fractonic topologically ordered state in \cite{Vijay2016phm1603.04442} 
	are encoded also in an algebraic variety, which is defined by the common zeros of a set of polynomials over a finite field. 
	
	In contrast, the algebraic variety  in our case is a way to organize the data of generalized {higher-moment or subdimensional polynomial global symmetry}
	or its gauge theory. The use of algebraic variety for our wide classes of theories do not require to be a gapped (fractonic) topological order.
         Our theories include gapless or gapped theories. 
         	
\subsection{From {Higher-Moment to Subdimensional or Subsystem Polynomial Global Symmetry}}

Let us relate the {algebraic (affine) variety and subvariety}
in \Sec{eq:AlgebraicVariety} to the patterns of polynomial in the {higher-moment or subdimensional or subsystem polynomial global symmetry}, or their gauge theories.
The studies of \emph{subdimensional or subsystem global symmetries} can be traced back to as early as \Refe{Batista2004scNussinov0410599,Nussinov2009zz0702377}
in condensed matter literature. 
Here we generalize the concept to study the  \emph{subdimensional or subsystem polynomial} global symmetry
For instance, subsystem global symmetry can act on 
lines \cite{YouDevakulBurnellSondhi2018oai1803.02369, Devakul2018fhz1808.05300} 
or planes \cite{Vijay2016phm1603.04442, YouDevakulBurnellSondhi2018zhj1805.09800, DevakulShirleyWang2019duj1910.01630},
for the bulk of 2+1D systems \cite{YouDevakulBurnellSondhi2018oai1803.02369, Devakul2018fhz1808.05300}  or 3+1D systems \cite{Vijay2016phm1603.04442, YouDevakulBurnellSondhi2018oai1803.02369, DevakulShirleyWang2019duj1910.01630} .

\begin{enumerate}[
label=\textcolor{blue}{\arabic*}., ref=\textcolor{blue}{\arabic*}]
\item 
\emph{From a degree-1 higher-moment symmetry to a degree-0 ordinary global symmetry in subdimensions}:
Recall the degree-1 polynomial global symmetry of \Eq{eq:degree-1-sym},
$\Phi\to\re^{\ii Q(x)}\Phi =   \re^{\ii (\Lambda_{i}  x_i + \Lambda_0)}\Phi$ acting on the matter field $\Phi$ on $ \R^n$.
 We can relate this degree-1 polynomial global symmetry to a
 degree-0 ordinary global symmetry by taking the constant surface solution of
$$(\Lambda_{i}  x_i + \Lambda_0)=\Lambda_{\text{constant}}$$
for a certain $(n-1)$D subdimensional space (e.g. plane) of $x_i \in  \R^n$.

\item \emph{From a degree-2 higher-moment symmetry to a degree-1 higher-moment or degree-0 ordinary global symmetry}:
Recall the degree-2 polynomial global symmetry of \Eq{eq:degree-2-sym},
$\Phi \to \re^{\ii Q(x)}\Phi =   \re^{\ii ( \Lambda_{i,j}  x_i x_j  + \Lambda_{i}  x_i + \Lambda_0) }\Phi$
acting on the matter field $\Phi$ on $ \R^n$.
 We can relate this degree-2 polynomial global symmetry to a
 degree-1 symmetry by restricting to an appropriate the constant $x_i= c_i$ space for some specific $x_i$.
 For example, we have
$$( \Lambda_{i,j}  x_i x_j  + \Lambda_{i}  x_i + \Lambda_0) \mid_{(x_i=c_i)}
=( \Lambda_{i,j}  c_i x_j  + \Lambda_{i}  c_i + \Lambda_0)$$
for a certain $(n-1)$D subdimensional space of $x_i \in  \R^n$. 
Moreover, we can reduce to a degree-0 ordinary global symmetry, if there is an intersecting subspace between
 the constant spaces of $x_i= c_i$ and $x_j= c_j$, e.g.
 $$( \Lambda_{i,j}  x_i x_j  + \Lambda_{i}  x_i + \Lambda_0) \mid_{({x_i=c_i}, {x_j=c_j})}
=( \Lambda_{i,j}  c_i c_j  + \Lambda_{i}  c_i + \Lambda_0).$$
Depend on the $ \Lambda_{i,j}$ and $\Lambda_{i}$, there could be a different constant surface by solving the polynomial with a different set of constraints.

For example, given a two-variable quadratic equation 
\bea \label{eq:degree-2-sym-conic-section}
Q(x) :=Q(x_1,x_2)=  \Lambda_{1,1}  (x_1)^2+ \Lambda_{1,2}  x_1 x_2 + \Lambda_{2,2}  (x_2)^2  + \Lambda_{1}  x_1+ \Lambda_{2}  x_2 + \Lambda_0,
\eea
we can solve the constant space to be an ellipse, a parabola, a hyperbola, also possibly a circle, a line, or two crossing lines, etc.
The solution is a quadratic algebraic curves through the well-known conic section.
In other words, if we apply the degree-2 polynomial global symmetry of
$\re^{\ii Q(x)}$ under \Eq{eq:degree-2-sym-conic-section},
we can find the 
degree-2 polynomial global symmetry on the $\R^n$ reduced to the ordinary degree-0 global symmetry on the
algebraic curves (an ellipse, a parabola or a hyperbola, etc) through the well-known conic section.

\item  \emph{From a degree-$(\tm-1)$ higher-moment symmetry to a subdimensional lower-degree (higher-moment or ordinary) global symmetry}:
Recall the general degree-$(\tm-1)$ polynomial global symmetry of 
	 \Eq{eq:degree-sym},
	$$
	\Phi\to\re^{\ii Q(x)}\Phi
	$$
	where $Q(x)$ is a polynomial with degree at most $(\tm-1)$, say
\Eq{eq:Qx-general}
$$
Q(x) :=	(\Lambda_{i_1,\dots,i_{\tm-1}}  x_{i_1} \dots x_{i_{\tm-1}} +\dots+ \Lambda_{i,j}  x_ix_j + \Lambda_{i}  x_i + \Lambda_0\big).
$$
We can reduce the degree-$(\tm-1)$ higher-moment symmetry in $\R^n$
to a lower degree-$(\tm-2)$ higher-moment symmetry in $\R^{n-1}$
by restricting to a specific subspace 
$$x_i= c_i.$$
\end{enumerate}
More generally, we can solve the polynomial with certain constraints as
a lower-degree polynomial. This is related to the concepts of  variety and subvariety
in \Sec{eq:AlgebraicVariety}, and the mathematical concepts of embedding and foliation of subspaces.
Indeed, the foliation concepts are powerful and applied recently in fracton literature, e.g. \cite{Shirley2017suz1712.05892, 2019arXiv190709048SShirleySlagleXieChen}.
It is also pointed out that the concept of \emph{spacetime embedding} may be treated as a quantum mechanical way as
a quantized excitation, named the \emph{embeddon}  \cite{Wang2019aiq1909.13879}. 
Therefore, it will be illuminating to
revisit all the above new \emph{gauge theories} of
{higher-moment or subdimensional polynomial global symmetry}
in a fully quantum mechanical set up in the future.

{\emph{Note added}: The sequel of this work as companions include \Refe{Wang2019aiq1909.13879} 
and \Refe{WXY3Wang2019mtt1912.13485}. 
} 

\section{Acknowledgements} 

JW thanks Meng Cheng and Trithep Devakul for helpful communications on References in the literature.\footnote{After
the completion of our work, we thank Meng Cheng for pointing out a potentially related  \Refe{GriffinPetrHorava2014bta1412.1046} and references therein on the study of the
polynomial shift symmetries.
This is a generalization to allow for an extension of the constant shift symmetry to  a polynomial shift symmetry in the spatial coordinates.
Although the essences of our and their ideas are related, the outcomes and motivations are dramatically different. We do not yet know the precise
correspondence between our results and theirs.}
JW is supported by
Center for Mathematical Sciences and Applications at Harvard University.
KX  is supported by Harvard Math Graduate Program and
``The Black Hole Initiative: Towards a Center for Interdisciplinary Research,'' Templeton Foundation. 
This work is also supported by 
NSF Grant DMS-1607871 ``Analysis, Geometry and Mathematical Physics'' 
and Center for Mathematical Sciences and Applications at Harvard University.

\bibliographystyle{Yang-Mills.bst}
\bibliography{fracton-embeddon-sigma.bib}

\end{document}